\documentclass[aps,preprint,showpacs]{revtex4}
\usepackage{graphics}
\usepackage{amssymb}
\usepackage{amsmath}
\usepackage{amsfonts}
\def\mmat#1{{\hbox{$\sf{#1}$}}}
\begin{document}
\title{Overlapping Resonances in the Resistance of Superposition States to
Decoherence}
\author{Asoka Biswas$^1$, Moshe Shapiro$^{2}$, and Paul Brumer$^1$}
\affiliation{$^1$Chemical Physics Theory Group, Department of
Chemistry, and Center of Quantum Information and Quantum Control,
University of Toronto, Toronto, Ontario M5S 3H6, Canada\\
$^2$Department of Chemistry,
University of British Columbia, Vancouver, V6T1Z1, Canada and \\
Chemical Physics Department, The Weizmann Institute of Science, Rehovot 76100, Israel}
\date{\today}
\begin{abstract}
Overlapping resonances are shown to provide new insights into the
extent of decoherence experienced by a system superposition state
in the regime of strong system-
environment coupling. As an example of
this general approach, a generic system comprising spin-half
particles interacting with a thermalized oscillator
environment is considered.
We find that (a) amongst the collection of parametrized
Hamiltonians, the larger the overlapping resonances contribution,
the greater the maximum possible purity, and (b) for a fixed
Hamiltonian, the larger the overlapping resonances contribution,
the larger the range of possible values of the purity as one
varies the phases in the system superposition states. Systems
displaying decoherence free subspaces show that largest
overlapping resonances contribution.
\end{abstract}
\pacs{03.65.Yz,03.67.Pp} \maketitle

\section{Introduction}

Coherent control of atomic and molecular processes is a 
subject of considerable ongoing 
interest\cite{brumer1,rice}. Specifically, a number of successful
control scenarios have been theoretically proposed and experimentally
implemented. The vast majority of well characterized scenarios
are, however, based upon  molecules treated in isolation, with the 
understanding that systems in an external environment are subject
to decoherence effects\cite{joos,shlosshauer} that can be deleterious to 
control\cite{spanner}.
Decoherence effects of this kind are also detrimental to ``quantum
technologies", such as quantum information processing and quantum
computing\cite{quantumcomputing}  insofar as they cause the loss of 
the essential quantum characteristics that make these research areas  
interesting.

As a consequence, efforts are ongoing to find suitable
mechanisms to reduce or compensate for decoherence effects.
For example, several
techniques, such as quantum dynamical decoupling (QDD) \cite{viola}
and the use of decoherence free subspace (DFS) \cite{lidar1}, have
been proposed in the quantum information community.
However, the QDD method is quite
challenging to implement using the present technology, and 
the DFS requires explicit symmetries in the Hamiltonian, which are
not generic. Hence, methods for identifying conditions for states that are
resistant to decoherence is of great interest.


Indeed, considerable effort has gone into identifying
characteristic features of particular system states, interacting
with different environments, that lead to stability against
decoherence. Sample approaches include the
general predictability sieve approach of Zurek {\emph
et al.} \cite{sieve}, and predictions of the unusual stability of
coherent matter states, low lying eigenstates, and localized
states \cite{zurek, arjendu} for typical environmental coupling.

In this paper we introduce a new perspective, based on overlapping resonances,
to understanding and optimize the
resistance of system superposition states to decoherence.
The approach  applies in the important
regime where system-environment coupling is large.
Specifically, we show that interference between the overlapping resonances
that exist in this regime can be used to build system superposition states
that are relatively resistant to decoherence. The extent to which such states are
stable against decoherence is shown to directly correlate with the
degree to which the system-environment Hamiltonian displays overlapping resonances.
Thus, even if the system does not possess symmetry properties such as
decoherence free subspaces (DFS), states can be constructed
which are more resistant to decoherence than are others.

From a formal viewpoint,
the approach described herein constitutes a major extension,
to product Hilbert
spaces characteristic of open quantum systems, of our work on
controlling radiationless transitions in large molecules
\cite{christo}. There, population transfer between a ${\cal
Q}$-subspace and the remainder of the Hilbert space, the ${\cal
P}$-subspace, was minimized by preparing initial superpositions
in the ${\cal Q}$-subspace, relying on the presence of
overlapping resonances.  This approach was applicable to the
case where the full Hilbert space was partitioned into the {\it
sum} of two subspaces. By contrast, decoherence involves the
effect of an environment on the system, i.e. {\it a
product} of two Hilbert spaces. The difference is substantial and
significant.

The inter-relationship between overlapping resonances and the ability to
design states that are resistant to decoherence is general. It proves
useful, however, to introduce the approach via a specific generic
example. For this reason we choose a particular class of spin-boson problems.
Note that system decoherence is a manifestation of the loss of quantum 
information when one neglects the environment with which the system is 
entangled\cite{shlosshauer}.  
As such, decoherence can occur effectively even with a small 
environment\cite{jacquod} as is manifest in the example below.

The structure of the paper is as follows. In Sec. II, we introduce
a generic system that describes a spin system interacting with a environment,
providing a specific example for this general approach.
Section III, considers the nature of decoherence in this system, and its
relationship to overlapping resonance. Some specific numerical examples
are provided in Sect. IV to
demonstrate correlations between overlapping resonances and
reduced state decoherence. Sect. V provides a summary.

\section{A Generic Class of Hamiltonians}

The spin-boson problem has been extensively studied in Chemistry in 
various scenarios and for a wide variety of applications\cite{spinboson}.
Here we consider,  as a specific
example, the control of decoherence in a simple spin system interacting
with a  small thermal environment, i.e.  a single spin-half system
and an environment comprising a  single-mode oscillator.  
The most general system-environment Hamiltonian for this case is given by
\begin{equation}
H/\hbar=\omega a^\dag
a+\frac{\omega_0}{2}S_z+g_r(S_+a+S_-a^\dag)+g_{nr}(S_+a^\dag+S_-a)+g_{ph}S_z(a+a^\dag)\;,
\label{hamil}
\end{equation}
where $a$ and $a^\dag$ are the environment annihilation and creation
operators, $\omega$ is the frequency of the environment mode,
$\hbar\omega_0$ is the energy difference between the two spin-half
states $|\pm\rangle$, and the spin operators are defined as
$S_\pm=|\pm\rangle\langle \mp|$,
$S_z=|+\rangle\langle+|~-~|-\rangle\langle -|$ . Note that the
rotating wave approximation (RWA) has not been invoked: the $g_r$
term in the Hamiltonian is the spin-environment interaction that is
retained under the RWA, whereas the $g_{nr}$ term is that  which
would be neglected in the RWA. The $g_{ph}$ term corresponds to
decoherence of the spin due to the operator $S_z$, which induces a
relative phase between the spin states $|+\rangle$ and
$|-\rangle$. The $g_r$ and $g_{nr}$ terms mediate energy exchange
between the spin and the environment. Our focus below is in the region of
strong coupling, where $g_{nr}$ and $g_r$ are both greater than
$\omega_0$.  It is important to emphasize that, although the
environment is small (a single oscillator) its coupling to the
system does induce significant system decoherence.

The class of Hamiltonians [Eq. (\ref{hamil})]  possesses a plane
defined by $g_r = g_{nr}$ in the three dimensional
$g_{nr},g_r,g_{ph}$ parameter space upon which the system displays
a decoherence free subspace (DFS). That is, in these special cases
the Hamiltonian Eq. (\ref{hamil}) leads to a DFS due to  a Lie
algebraic symmetry. In these cases, the interaction part of the
total Hamiltonian can be written as ${\cal E}_I\otimes B_I$, where
$B_I=(a+a^\dag)$ is the environment operator, and the spin operator
${\cal E}_I$ can be written in a closed matrix form in the
$(|+\rangle,|-\rangle)$ spin basis:
\begin{equation}
\label{error}{\cal E}_I=\left(\begin{array}{cc}g_{ph}& g \\ g &
-g_{ph}\end{array}\right)\;,\;g=g_r=g_{nr}\;,
\end{equation}
The DFS
corresponds to the eigenstates of the ${\cal E}_I$ matrix. For
other values of $g_{nr}$ and $g_r$, of primary interest here,
the system does not have a DFS.


\section{Decoherence of a Spin Superposition}

We consider the stability against decoherence of 
an initial spin superposition state $|\psi\rangle =c_+|+\rangle+
c_-|-\rangle$, where $c_\pm$ are complex amplitudes and the environment
is in thermal equilibrium at temperature $T$. The latter is
described in terms of the oscillator eigenstates $|n\rangle$ by
the environment density matrix $\rho_B=\sum_np_n|n\rangle\langle n|$,
$p_n=e^{-E_n/k_BT}/(\sum_n e^{-E_n/k_BT})$ and
$E_n=\left(n+1/2\right)\hbar\omega$. In order to follow the
temporal evolution of the system we expand the total (system+environment)
evolution operator $U$ in terms of $|\gamma\rangle,$ the
eigenstates of the full Hamiltonian [i.e., $(E_\gamma-H)
|\gamma\rangle=0$] with $U=\sum_\gamma e^{-iE_\gamma
t/\hbar}|\gamma\rangle\langle \gamma |$. Expanding the eigenstates
$|\gamma\rangle$ in the zeroth order basis of ``bare" spin+environment
states $\{|\pm,n\rangle\equiv|\pm\rangle\otimes|n\rangle\},$ (i.e.
the eigenstates of $H$ without the spin-environment interaction) gives,
for the time-evolved density matrix of the total spin+environment system:
\begin{equation}
\rho(t)=\sum_{n}p_nU\left[|\psi\rangle\langle \psi |\otimes
|n\rangle\langle n|
\right]U^\dag=\sum_{\gamma,\gamma',n}p_ne^{-i(E_\gamma -
E_{\gamma'})t}|\gamma\rangle\langle \gamma'|\langle
\gamma|\psi,n\rangle\langle \psi, n|\gamma'\rangle\;.
\end{equation}
Further, tracing over the environment states, gives the following reduced
system density matrix elements $(\rho_s)_{kl}$ in the basis of
spin states $|k\rangle$ and $|l\rangle$ where $k,l$ denote either the
plus or minus state:
\begin{equation}
(\rho_s)_{kl}=\sum_{m,n,\gamma,\gamma'}p_ne^{-i(E_\gamma-E_{\gamma'})t}\langle
k,m|\gamma\rangle\langle\gamma|\psi,n\rangle\langle
\psi,n|\gamma'\rangle\langle\gamma'|l,m\rangle\;.
\end{equation}
Expanding $|\psi\rangle$ as $c_+ |+\rangle + c_- |-\rangle$, 
we can rewrite the above expression as
\begin{equation}
(\rho_s)_{kl}=|c_+|^2Q_{kl}(t)+|c_-|^2
R_{kl}(t)+c_+c_-^*P_{kl}(t)+c_-c_+^*T_{kl}(t)\;,
\label{timeevolution}
\end{equation}
where ${\mmat P}$,${\mmat Q}$,${\mmat R}$,${\mmat T}$ are complex
matrices with elements
\begin{eqnarray}
P_{kl}(t) & = &
\sum_{m,n,\gamma,\gamma'}p_ne^{-i(E_\gamma-E_{\gamma'})t}\langle
k,m|\gamma\rangle\langle\gamma'|l,m\rangle\langle\gamma|+,n\rangle\langle
-,n|\gamma'\rangle\;,\nonumber\\
Q_{kl}(t) & = &
\sum_{m,n,\gamma,\gamma'}p_ne^{-i(E_\gamma-E_{\gamma'})t}\langle
k,m|\gamma\rangle\langle\gamma'|l,m\rangle\langle\gamma|+,n\rangle\langle
+,n|\gamma'\rangle\;,\nonumber\\
R_{kl}(t) & = &
\sum_{m,n,\gamma,\gamma'}p_ne^{-i(E_\gamma-E_{\gamma'})t}\langle
k,m|\gamma\rangle\langle\gamma'|l,m\rangle\langle\gamma|-,n\rangle\langle
-,n|\gamma'\rangle\;,\nonumber\\
T_{kl}(t) & = &
\sum_{m,n,\gamma,\gamma'}p_ne^{-i(E_\gamma-E_{\gamma'})t}\langle
k,m|\gamma\rangle\langle\gamma'|l,m\rangle\langle\gamma|-,n\rangle\langle
+,n|\gamma'\rangle\,\label{qrst}
\end{eqnarray}

Consider the decoherence of the spin state, quantified in terms of
the purity \cite{xupei} $S$ of the spin, where  $S={\rm
Tr}(\rho_s^2)$. Given the above results, $S$ can be evaluated via
\begin{equation}
S=\sum_{k,l\in +,-}\langle k|\rho_s|l\rangle\langle
l|\rho_s|k\rangle=\sum_{k,l\in +,-}|(\rho_s)_{kl}|^2\,
\label{purity}
\end{equation}
so that, using $P_{kl}^*=T_{lk}$:
\begin{eqnarray}
&&S =  |c_+|^4\sum_{k,l\in +,-}|Q_{kl}|^2+|c_-|^4 \sum_{k,l\in
+,-}|R_{kl}|^2 \nonumber\\
&&+|c_+|^2|c_-|^2\sum_{k,l\in +,-}\left[|P_{kl}|^2+|T_{kl}|^2+\{Q_{kl}R_{kl}^*+Q_{kl}^*R_{kl}\}\right]\nonumber\\
&&+2 {\rm Re} \left[\sum_{k,l\in +,-}\left[c_+^2c_-^{*2}P_{kl}P_{lk}
+|c_+|^2c_+^*c_-[Q_{kl}P_{kl}^*+P_{lk}^*Q_{kl}^*]
+|c_-|^2c_+^*c_-[R_{kl}P_{kl}^*+P_{lk}^*R_{kl}^*]\right]\right] . \label{mix}
\end{eqnarray}
The first three of these terms depend solely on the magnitude of
the $c_{\pm}$ coefficients, whereas the remainder of the terms
depend upon the phases of the coefficients as well.  The latter
are clearly contributions to $S$ that rely upon quantum
interference between the spin states.

Note, for comparison later below, that $S=1$ denotes a pure state. 
If the state is completely
decohered in the spin basis then $(\rho_s)_{kl} = \frac{1}{2}\delta_{kl}$ 
so that the purity, by Eq. (\ref{purity}), is $S=2(1/2)^2 = 1/2$.

With the $\mmat{P,Q,R,T}$ matrices known, one can optimize $S$
with respect to $c_+$ and $c_-$ to determine the state with
maximum purity $S$ at a fixed time. In particular,  the relative
phase of $c_+$ and $c_-$ is  important if either $Q_{kl}$ or
$R_{kl}$ is non-zero, and $P_{kl}$ is non-zero .
Although not evident from Eq. (\ref{mix}), numerical results below
clearly demonstrate a reliance on this relative phase to maximize
$S$.

\section{Overlapping Resonances and System Purity}

\subsection{Overlapping Resonances}

As noted above, the bare states $|\pm,n\rangle$ 
are not eigenstates of $H$, and  will
evolve in  the presence of the system-environment coupling.  
Insight into the nature of this time evolution is afforded by expanding the 
zeroth order state in the exact eigenstates $|\gamma\rangle$. 
For example, a bare state such as $|+,j\rangle$ expands as
\begin{equation}
|+,j\rangle = \sum_{\gamma} |\gamma\rangle \langle\gamma|+,j\rangle,
\end{equation}
and evolves  as 
\begin{equation}
|+,j\rangle_t = \sum_{\gamma} |\gamma\rangle \langle\gamma|+,j\rangle
e^{-i E_{\gamma}t/\hbar}
\end{equation}
The $E_\gamma$ dependence of the square of the expansion coefficient, i.e. 
$|D_\gamma|^2 \equiv |\langle\gamma|+,j\rangle|^2$ 
provides the energy width
over which the zeroth order state $|+,j\rangle$ is spread due to the
system-environment interaction. In the simplest cases 
the inverse of this width provides a 
qualitative measure of the time scale for the evolution of $|+,j\rangle$.
Hence, the zero order states are indeed resonances with a characteristic
width $|D_\gamma|^2$. This is the analog, in a bound state spectrum, of the 
well known resonance in the continuum.

Similarly, by analogy to the continuum case, we can define 
{\it overlapping resonances}, as
resonances that overlap in energy space, i.e. resonances that share a common
$|\gamma\rangle$ in their respective eigenstate expansions. 
These are the precise bound state analogues to
overlapping resonances in the continuum [e.g.  
Fig 1, Ref. \cite{frishmanshapiro}].
Examples of bound state overlapping resonances in molecular
systems have also been previously reported\cite{christo}.

\begin{figure}[htbp]				        
\begin{center}
\scalebox{0.6}{\includegraphics{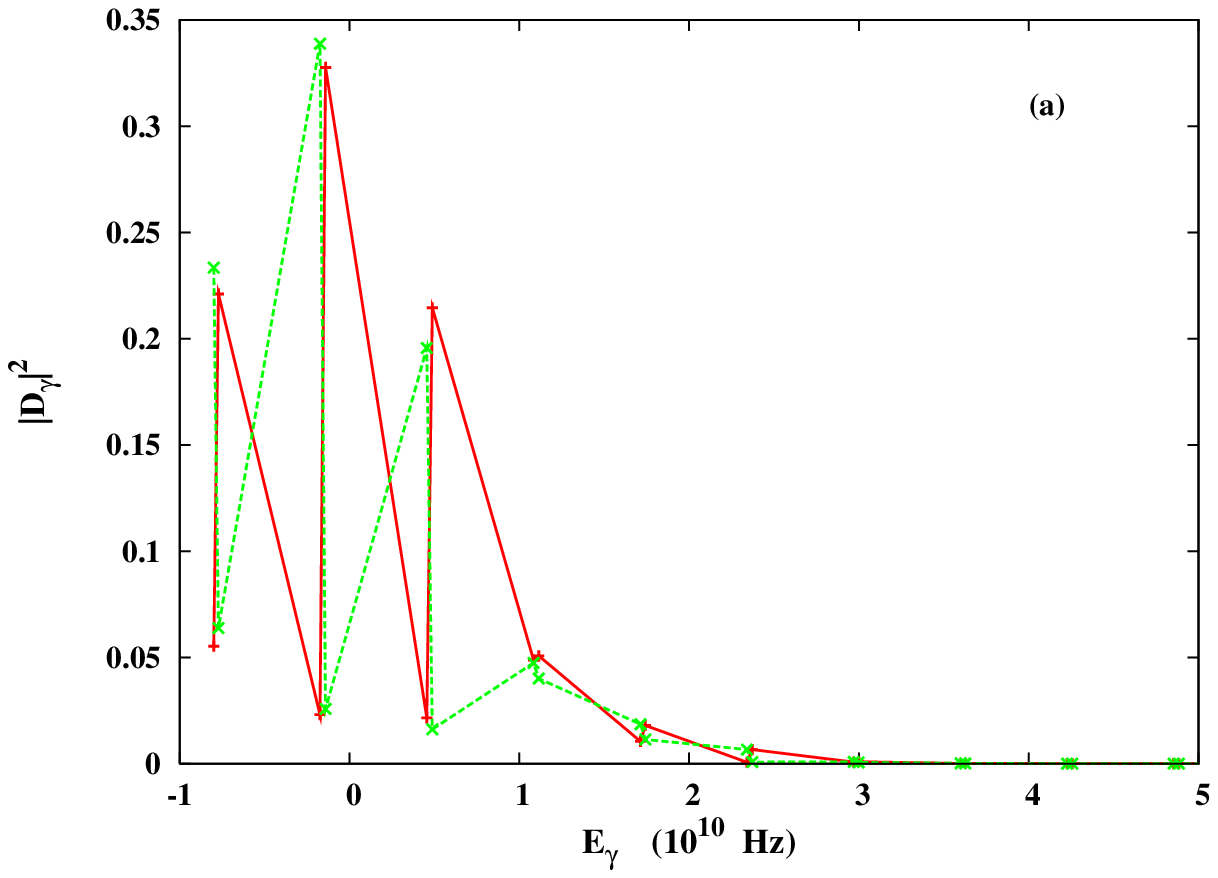}} \\
\scalebox{0.6}{\includegraphics{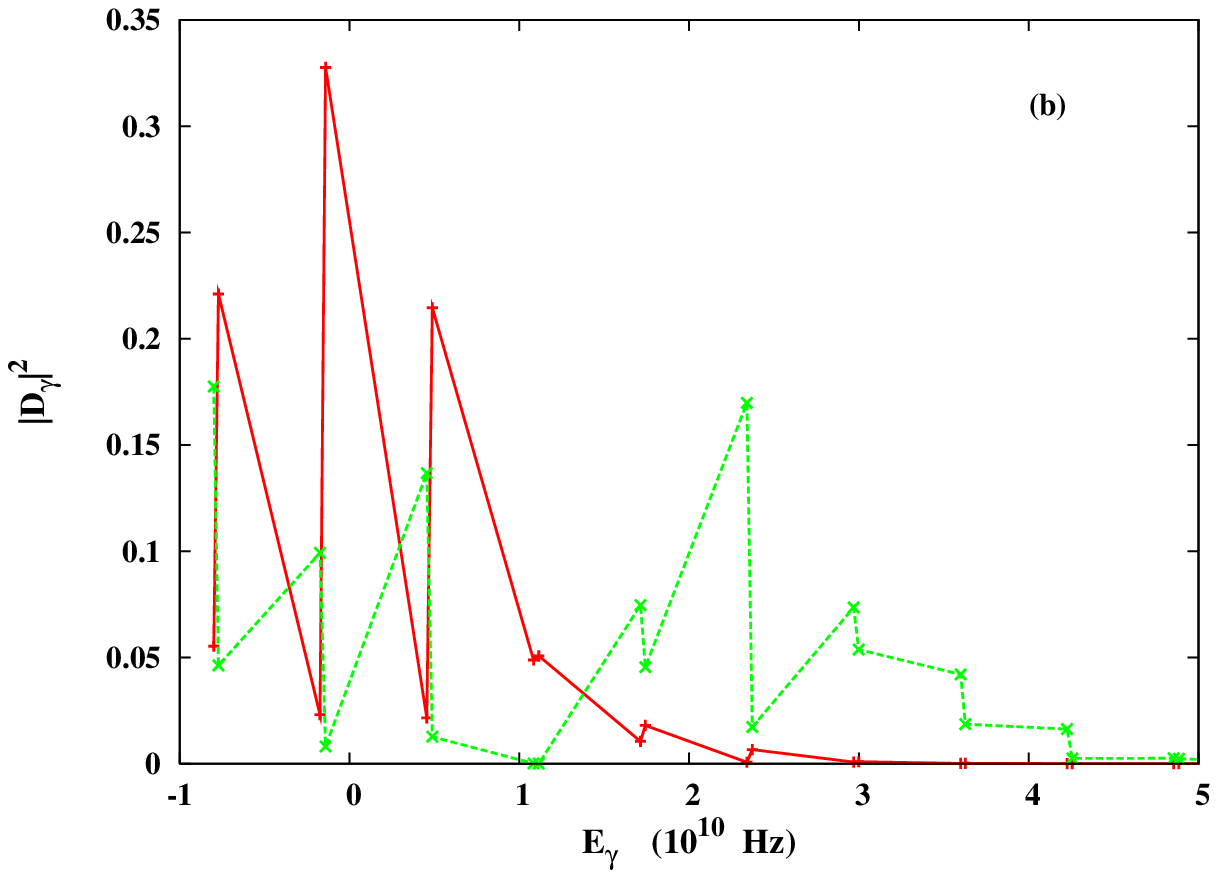}} \\
\scalebox{0.6}{\includegraphics{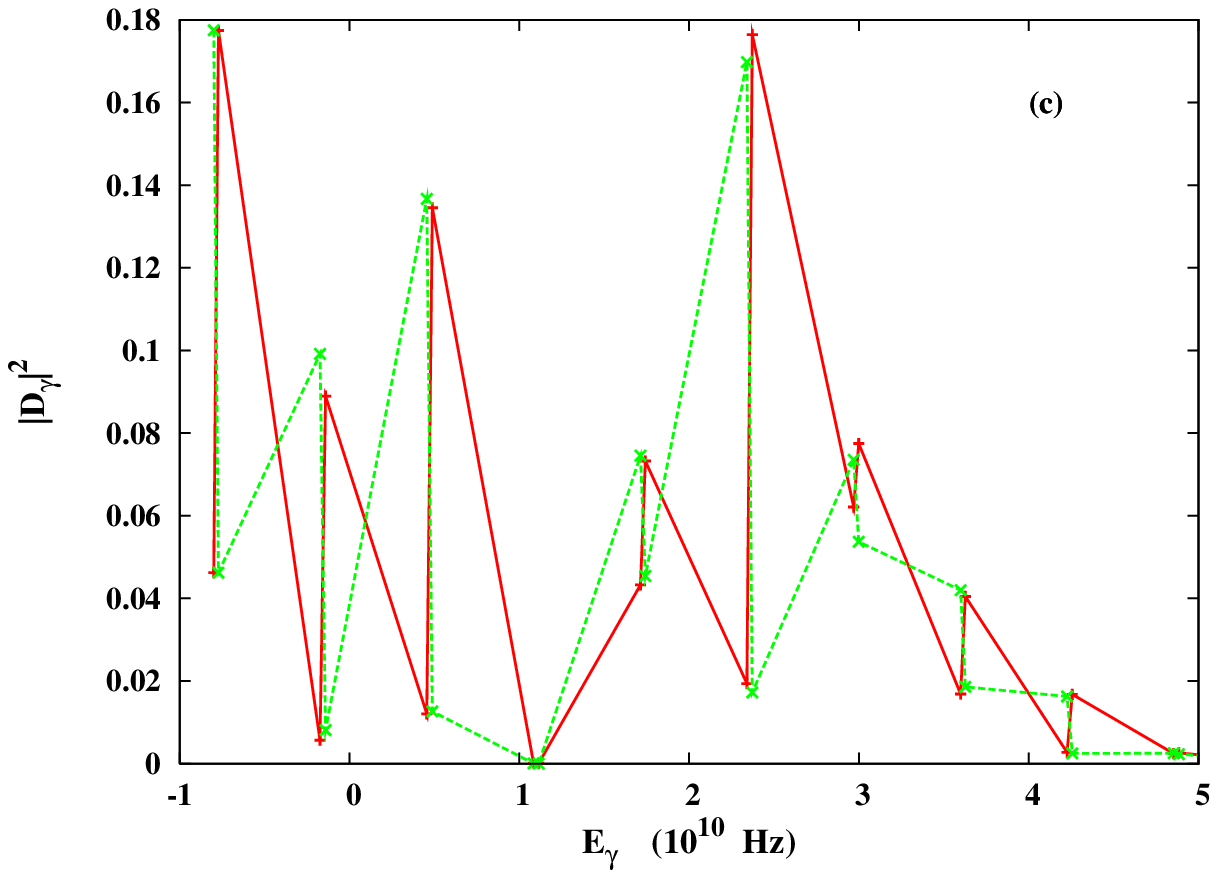}} \\
\caption{Plot of the overlap $|D_\gamma|^2$ between various bare
states and the eigenstates $|\gamma\rangle$ as a function of
the energy
$E_\gamma$ : (a) $|+,0\rangle$ ,
$|-,0\rangle$, (b) $|+,0\rangle$, $|-,2\rangle$, 
(c) $|+,2\rangle$,  $|-,2\rangle$. The data points for the
first of each pair are connected by a solid line whereas
the data points for the second of each pair are connected
by solid lines. Data are for 
the parameters $\omega=2\pi$ GHz, $\omega_0=2\pi \times 100$ MHz, 
$g_r=g_{nr}=2 \pi$ GHz and $g_{ph}=0$.}
\label{fig:overlapping}
\end{center}\end{figure}

In the case of the spin-boson Hamiltonians
given by Eq. (\ref{hamil}), overlapping resonances were clearly
evident in, but not limited to, the strong coupling regime
characterized by $g_r$ and $g_{nr}$ greater than $\omega_0$. 
They  occur as
two states $|k,m\rangle$ and $|l,n\rangle$ with $k \ne l$
and/or $m \ne n$ ($k,l\in +,-$) that have non-zero overlap with the
same eigenstate $|\gamma\rangle$. 
Examples of such overlapping resonances are shown in 
Fig. \ref{fig:overlapping}. 
Quite clearly the overlap between the resonances
can be significant, enhanced by the fact that the overall width of each
resonances can be significantly larger than the energy spacing between the 
zeroth order states. The resonances are also seen to be rather 
highly structured.

\subsection{Contributions to {\rm Tr} $\rho_s^2$}

Consider then the role of these overlapping resonances in the dynamical
evolution of $\rho_s(t)$ given in Eq. (\ref{timeevolution}). 
Specifically, note
that all the matrices ${\mmat P}$,${\mmat Q}$,${\mmat R}$,${\mmat T}$
depend on overlapping resonance contributions.  Their dependences
differ, however, in the important case where
$k\ne l, k,l\in +,-$. In that case overlapping
resonances contribute to $T_{kl}$ and $P_{kl}$, but not to
$Q_{kl}$ and $R_{kl}$. That is, the terms in $(\rho_s)_{kl}$
that are sensitive to the relative phase of $c_+$ and $c_-$
[see Eq. (\ref{timeevolution})], and are hence
associated with phase control over the dynamics, are directly dependent on
overlapping resonances contributions.

>From Eq. (\ref{mix}), it is apparent that the purity is strongly
affected by overlapping resonances through the matrix elements
$Q_{kl}$, $R_{kl}$, $P_{kl}$, and $T_{kl}$. To quantify this
dependence, we explicitly identify overlapping resonances contributions
associated with
the plus and minus spin states. That is, we define
\begin{equation}
\label{overlap}A_{+-}=\sum_{m \ne n}A_+^{m,n}A_-^{m,n}\;,\;A_{\pm}^{m,n}=\left[\sum_\gamma
\left|\langle \pm,m|\gamma\rangle\langle
\gamma|\pm,n\rangle\right|\right]\;.
\end{equation}
For a pair of the oscillator states $|m\rangle$ and $|n\rangle$,
$A_+^{m,n}$ determines the overlap (summed over all eigenstates
$|\gamma\rangle$) between the bare states $|+,m\rangle$ and
$|+,n\rangle$, when the spin state is $|+\rangle$. If, for the
same pair of $|m\rangle$ and $|n\rangle$,  there exists
non-zero overlap $A_-^{m,n}$ between the bare state $|-,m\rangle$
and $|-,n\rangle$,  then $A_{+-}$  provides a measure
of the overlap between states $|+\rangle$ and $|-\rangle$,
that arises through the overlap between states
$|\pm,m\rangle$ and $|\pm,n\rangle$. Note that Eq. (\ref{overlap})
is somewhat similar to the off diagonal $T_{+-}$ and $P_{+-}$ terms at
$t=0$.
\begin{figure}
\begin{tabular}{cc}
\scalebox{0.6}{\includegraphics{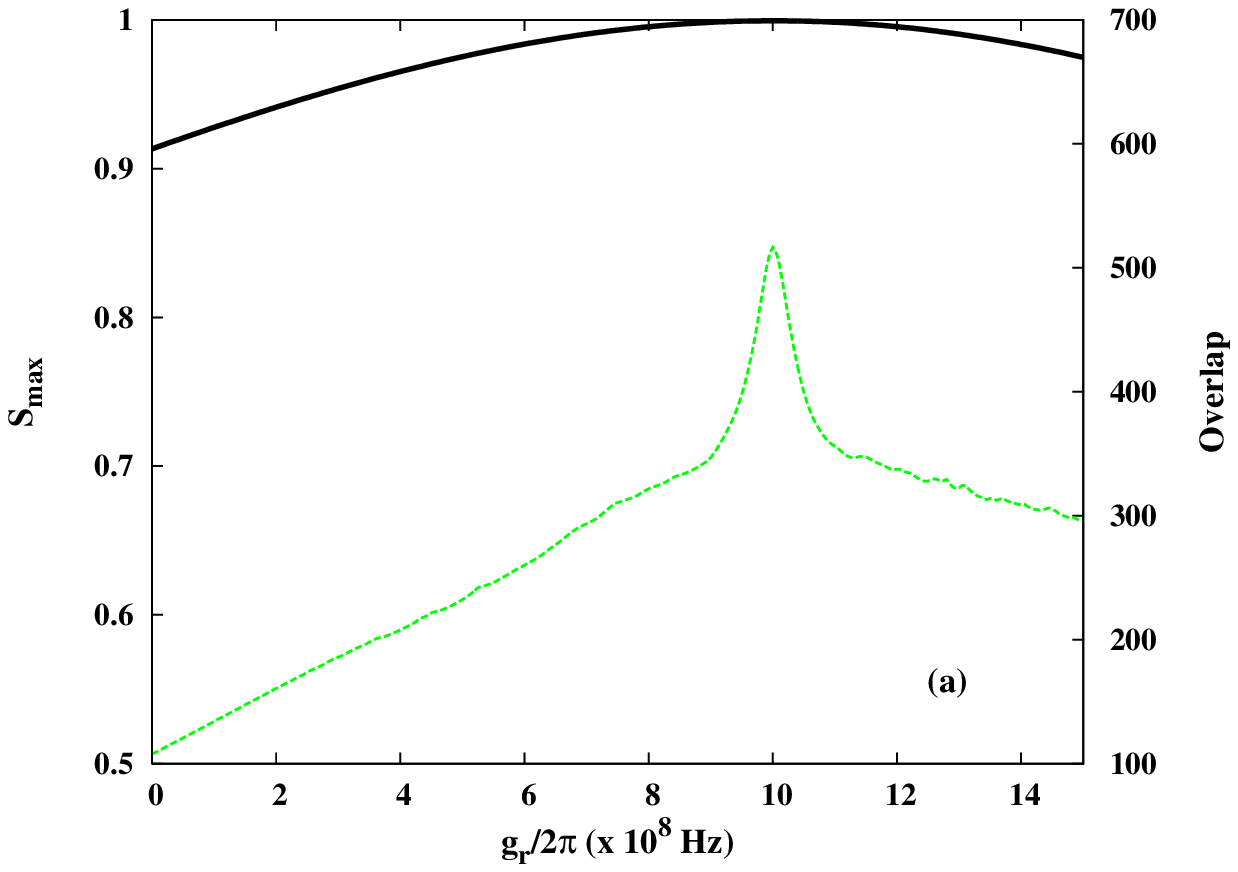}} &
\scalebox{0.6}{\includegraphics{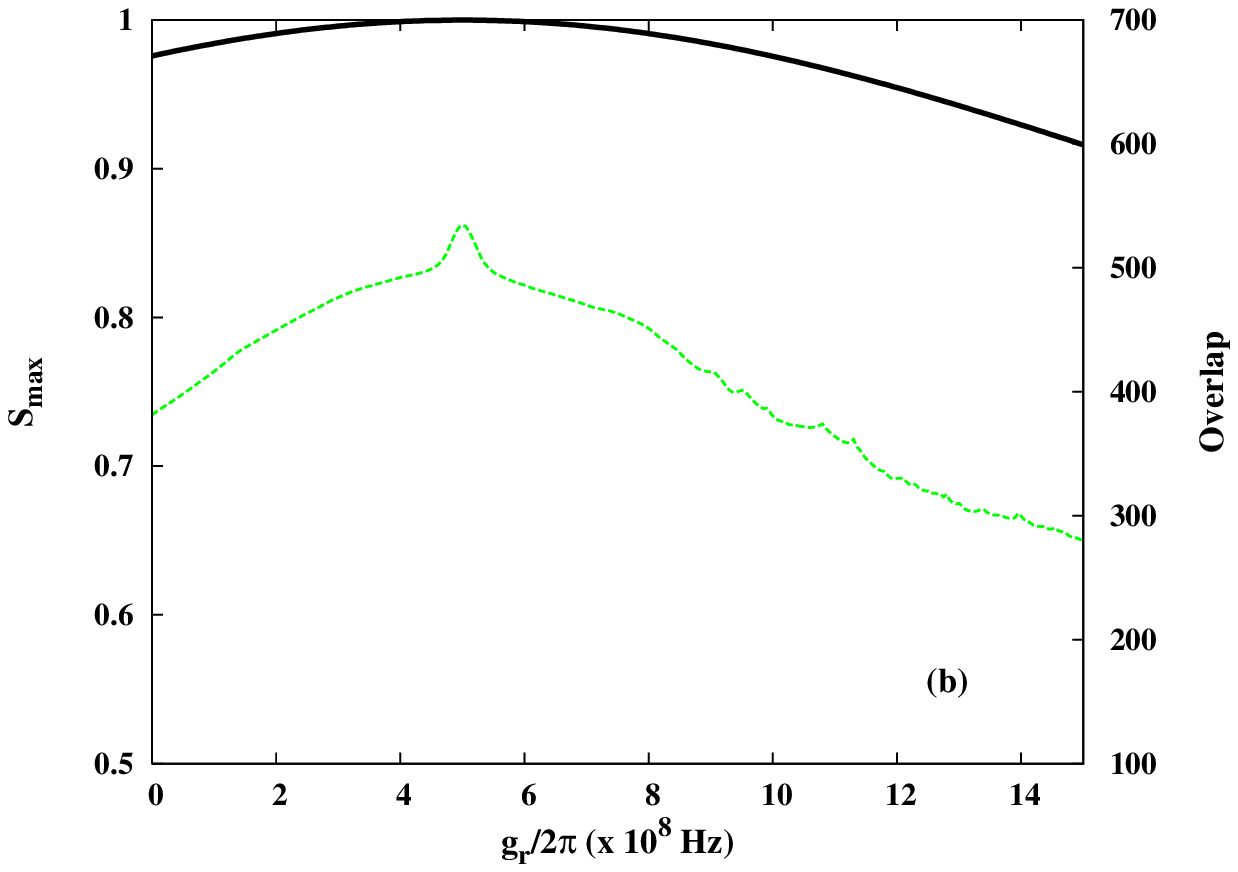}}
\end{tabular}
\caption{ The dependence of the maximum achievable
purity $S_{\rm max}$ of the system (solid line) and the
overlapping resonance $A_{+-}$ (dashed line; right y-axis)
on the coupling strength $g_r/(2\pi)$. Here (a) $g_{ph}=
2\pi\times 500$ MHz, $g_{nr}=2\pi$ GHz and (b) $g_{ph}= 2\pi$ GHz,
$g_{nr}=2\pi\times 500$ MHz. The remaining parameters are $\omega=2\pi
$ GHz, $\omega_0=2\pi\times 100$ MHz, and the temperature $T=25$
mK. The purity is calculated at a time $t=0.1$ ns.}
\label{oldfigure1}
\end{figure}

Note that the number and character of the overlapping resonances
are functions of the parameters in the total Hamiltonian, e.g.,
here $g_r$, $g_{nr}$, $g_{ph}$, $\omega$, and $\omega_0$. For
example, in the extreme case where there is no system-environment
coupling (i.e. $g_r=g_{nr}=g_{ph}=0$), the total Hamiltonian is
diagonal in the spin-boson bare state basis. The overlap between
the spin-boson bare states then vanishes, and in turn, so does the
overlap between the states $|\pm\rangle$. That is, as expected, in
absence the system of coupling to the environment, the spin states are
not broadened and thus do not overlap.

\subsection{Computational Results}

Consider then the relationship between overlapping resonances as
reflected in $A_{+-}$ and decoherence as embodied in the system
purity $S$. As an example, we examine dynamics where the environment
temperature T = 25 mK and choose to vary $g_r$, keeping the other
Hamiltonian parameters fixed. 

Numerical results are obtained by diagonalizing the Hamiltonian matrix
in the zeroth order basis of bare states $|\pm,n\rangle$. At the 
temperature of $T=25$ mK, a total of 20 oscillator states contribute,
allowing us to diagonalize a 40 $\times$ 40 dimensional Hamiltonian
matrix to obtain the desired $E_{\gamma}, |\gamma\rangle$.

To assess the decoherence we find,
for each value of $g_r$,  the maximum possible purity $S_{\rm
max}$ in the system at a fixed time $t$ by optimizing the
complex coefficients $c_\pm$. Results are shown in Fig. \ref{oldfigure1} where
both $S_{\rm max}$ and the overlapping resonance measure $A_{+-}$
are plotted as a function of $g_r$ at $t=0.1$ ns. The
correspondence between the behavior of these two functions is
evident, with larger $A_{+-}$ correlating well with larger $S_{\rm
max}$. That is, the greater the overlapping resonances
contribution, the more resistant is the optimal superposition to
decoherence. Note, significantly, that the optimally resistant
state need not be, and is often not, an eigenstate of the spin
Hamiltonian, but is rather a superposition of $|+\rangle$ and
$|-\rangle$ eigenstates.

Prominent in the graph (and some presented  below) is the DFS
point $g_r = g_{nr}$, where the $A_{+-}$ curve displays a sharp
peak, commensurate with $S_{\rm max}$ reaching its maximum value
of unity. That is, maximal purity  clearly correlates with the
maximum $A_{+-}$. The deviation of $S_{\rm max}$ from unity as one
moves away from this point is significant, particularly in
scenarios such as quantum computation, where extremely high
degrees of coherence are necessary.

The results clearly show that states take advantage of
overlapping resonances (and hence of quantum interference) to
increase the state purity. This is the case even if the
system-environment coupling is very large, as is clear from 
Fig. \ref{oldfigure1}(b),
where, for example, $g_r\sim 2\pi\times 1.5$ GHz ($g_r\gg
\omega_0$), but where the purity remains as large as 0.92 for
$g_{ph}=2\pi$ GHz.

\begin{figure}
\scalebox{0.6}{\includegraphics{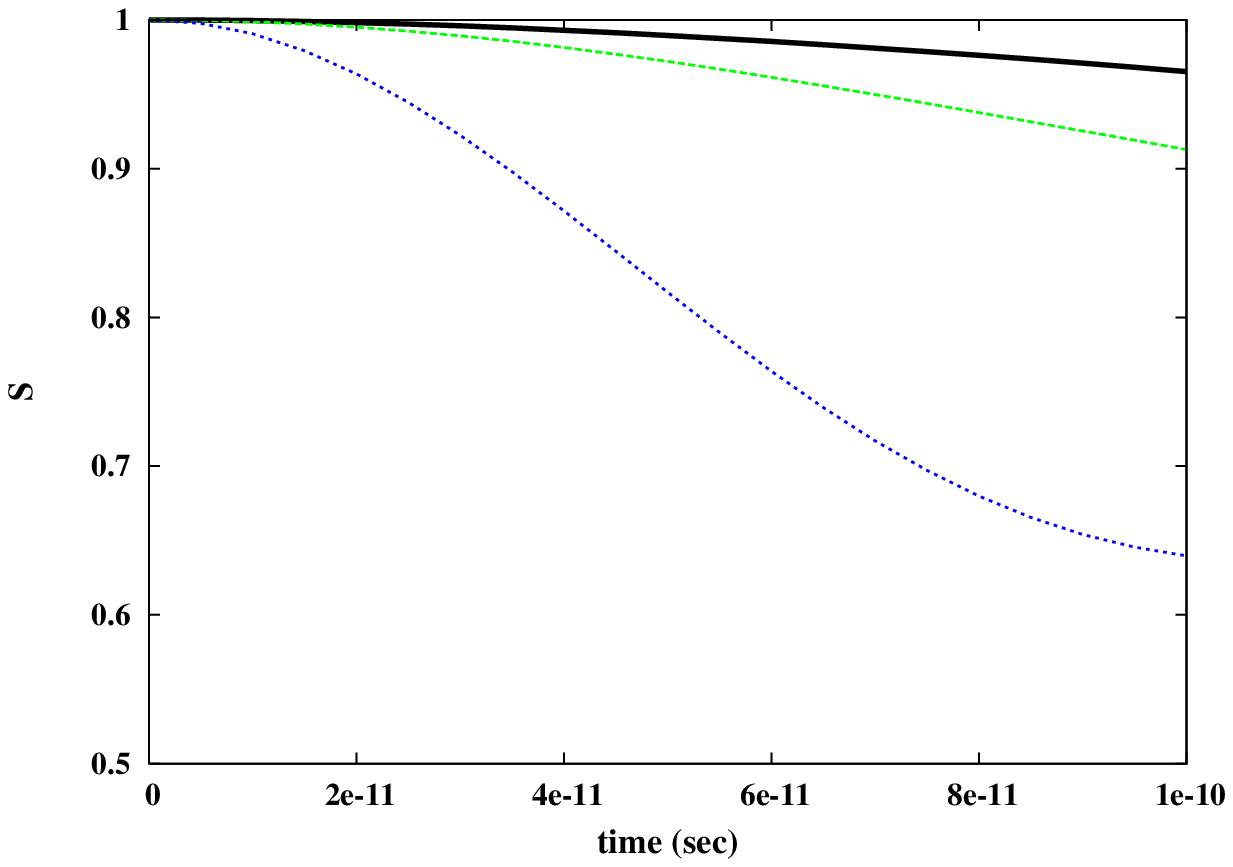}} \caption{
Variation of purity with time for the initial states
$(-0.539998|+\rangle+0.841665|-\rangle)$ (solid line),
$(|+\rangle+|-\rangle)/\sqrt{2}$ (dashed line), and
$|+\rangle$ (dotted line). Here the interaction strengths
$g_r=2\pi\times 400$ MHz, $g_{nr}=2\pi$ GHz, and
$g_{ph}=2\pi\times 500$ MHz. The other parameters are as in Fig.
\ref{oldfigure1}.}
\label{oldfigure2}
\end{figure}

\begin{figure}
\scalebox{0.6}{\includegraphics{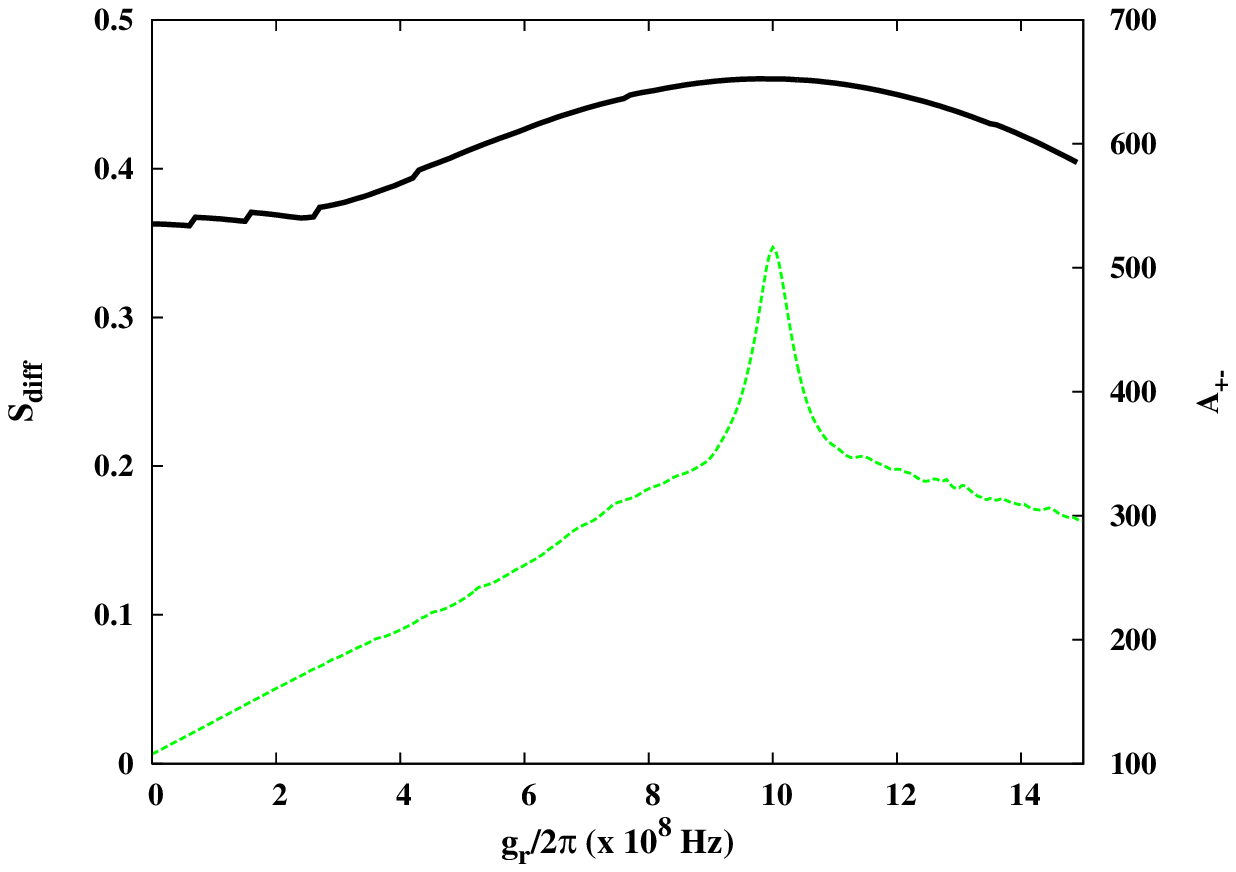}} \caption{Variation of
$S_{\rm diff}$ (solid line) and the overlap $A_{+-}$ (dashed
line; right y-axis) with $g_r$ for $g_{nr}=2\pi$ GHz and
$g_{ph}=2\pi\times 500$ MHz. The remaining parameters are as
in Fig. \ref{oldfigure1}.}
\label{oldfigure3}
\end{figure}

The presence of overlapping resonances, as seen above, influences
the degree of decoherence when preparing initial states with
different initial coefficients $c_\pm$. Sample dependences of $S$
on $c_\pm$ are shown in Fig. \ref{oldfigure2} for the case of $g_r=2\pi\times
400$ MHz, $g_{nr}=2\pi$ GHz, and $g_{ph}=2\pi\times 500$ MHz [i.e.
parameters associated with Fig. \ref{oldfigure1}(a)]. Shown are the time
evolution of $S$ for three states, one being the best case
superposition $c_+=-0.539998$ and $c_-=\sqrt{1-|c_+|^2}$ which
shows $S_{\rm max} =0.96$ at the time $t=0.1$ ns where $S$ is
optimized. Comparison is made with the time evolution of $S$ for
two other sample states with different $c_\pm$ values: $c_+=1$ and
$c_+=c_-=1/\sqrt{2}$. The range of $S$ as a function of initial
state is large, with the $c_+ =1$ case showing almost fully mixed
behavior at long time. 
Interestingly, the curves shown do not cross
as a function of time. Hence the optimal superposition, determined
by coefficients obtained for the $t=0.1$ ns case is, in fact,
the optimal superposition for the times preceding that target time
as well. This was also confirmed by optimizations independently
carried out at shorter times.

Additional results showing how overlapping
resonances between spin states is manifest in decoherence is shown in 
Fig. \ref{oldfigure3}. Here,
for a given value of $g_r$, we obtain the $|c_+|$ and $|c_-|$ that
maximize $S$ and then vary the relative phase $\theta$ between
$c_+$ and $c_-$ to obtain $S_{\rm diff}$, defined  as the
difference between the maximum and minimum purity so attained.
This difference is plotted in Fig. \ref{oldfigure3} as a function of $g_r$, along
with  the corresponding value of the overlapping resonance
contribution $A_{+-}$. The extent to which $S_{\rm diff}$ varies
is seen to reflect the variation of $A_{+-}$. Specifically, as
$A_{+-}$ increases, so does the extent to which the purity varies
with $\theta$. Hence, in the presence of large overlapping
resonance, one can {\it actively\/} control the purity of the
system by taking advantage of the phase dependent quantum
interference contribution to $S$. Indeed, in accord with Eq.
(\ref{mix}), an observed variation of the decoherence with the
phase of a superposition state provides evidence of the presence
of overlapping resonances.

\begin{figure}
\scalebox{0.6}{\includegraphics{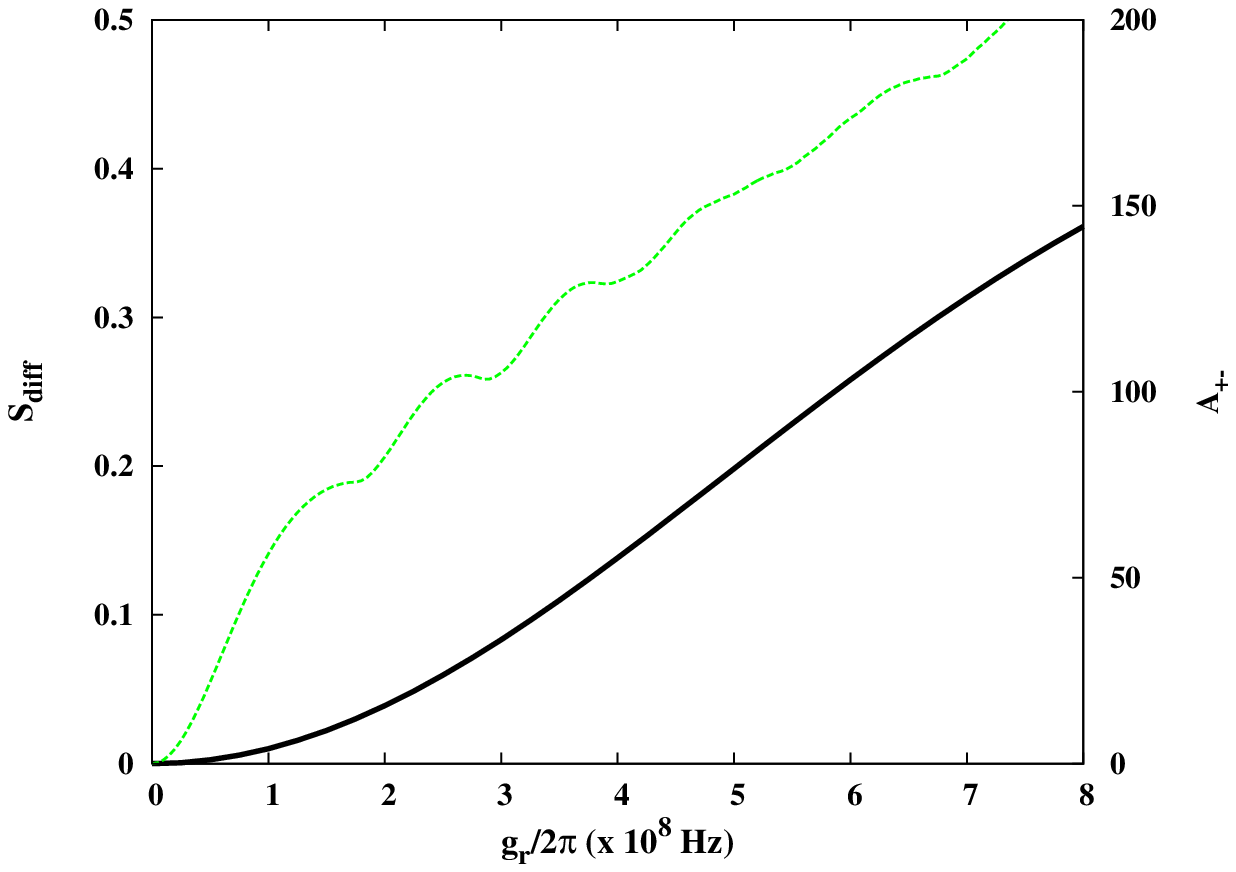}} \caption{Variation of
$S_{\rm diff}$ (solid line) and the overlap $A_{+-}$ (dashed
line; right y-axis) with $g_r=g_{nr}$ for $g_{ph}=0$.
The other parameters are as in Figs.\ref{oldfigure1}.}
\label{oldfigure4}
\end{figure}

The extent to which $S_{\rm diff}$ varies with $A_{+-}$ depends
upon the Hamiltonian parameters.  Consider, for example, an
interesting case of the nanoscale Hamiltonian system 
in Eq. (\ref{hamil}), the
interaction of a Cooper pair box  (CPB) with a nanomechanical
resonator \cite{namr}. This system is given by:
\begin{equation}
\label{namr}H_{CPB}=4E_C\delta nS_z-(E_J/2)S_x+\hbar\omega a^\dag
a+\hbar gS_z(a+a^\dag)\;,
\end{equation}
where $E_C$ and $E_J$ are the charging energy and the Josephson
energy of the CPB, $\omega$ is the fundamental frequency of the
resonator, and $\delta n$ lies between $-1/2$ and $1/2$. At the
degeneracy point $\delta n=0$ this Hamiltonian can be written in
the form of that of Eq. (\ref{hamil}) under a similarity
transformation, where we identify $\omega_0=E_J$, $g_{ph}=0$, and
$g_r=g_{nr}=g$. The similarity transformation is given by the
operator $e^{iS_y}$, with $S_y$ being the $y$-projection of the
spin operator.

>From the symmetry of the Hamiltonian (\ref{namr}), one finds a DFS
corresponding to the state $c_+=c_-=1/\sqrt{2}$ and
$c_+=-c_-=1/\sqrt{2}$. Hence, the value of $S_{\rm max}$ will
always be unity for all values of $g$, irrespective of the values
of $A_{+-}$. Nonetheless, overlapping resonances affect the
decoherence in this system. Specifically, as $A_{+-}$ increases
with $g$, the variation of $S$ with changes in the  relative phase
$\theta$ between $c_\pm$ also increases. This is shown in Fig. \ref{oldfigure4}
where  the difference $S_{\rm diff}$ between the maximum and
minimum values of $S$ and the values of $A_{+-}$ with $g$ are seen
to be strongly correlated.

On the other hand, if $g_r\ne g_{nr}$, there is no DFS in the
system.   In that case, the results (not shown) behave similarly to,
e.g., Fig. \ref{oldfigure1}.  That is,   a plot of $S_\textrm{max}$ and
$A_{\pm}$ as a function of $g_r$, with $g_r \ne g_{nr}$ and $g_{ph}=0$,
shows a clear correlation between the maximum achievable value of $S$ and
the overlapping resonance contribution.

\section{Summary}

We have introduced a new unifying approach to considering the
resistance to decoherence of system superposition states.
Specifically, using the generic class of spin-boson systems as an
example we have shown that overlapping resonances of the system
bare states, and the associated interference between these states,
provide insight into the dependence of the decoherence of system
superposition states in the important regime of strong system-environment
coupling. For example, within a given class of Hamiltonians, increasing
overlapping resonances contributions allow for states of higher
purity. The approach is completely general, although it has been
applied here to a spin 1/2 particle interacting with a single
thermal oscillator.  Extensions to other cases 
are underway. In the first
instance we anticipate that increasing the environment size will increase
the participation and effectiveness of overlapping resonances in
decoherence control.

{\it Acknowledgments}: This work was supported by NSERC and by the Centre
for Quantum Information and Quantum Control, University of Toronto.

\end{document}